\documentclass[12pt,a4paper]{article}%
\pdfoutput=1
\usepackage{amsmath,amssymb,amsfonts}
\usepackage[pdftex]{graphicx}
\usepackage{color}
\usepackage{float}
\usepackage[bf,footnotesize]{caption2}
\usepackage[]{hyperref}%
\numberwithin{equation}{section}
\setcaptionmargin{1cm}
\newcommand{\hhref}[1]{\href{http://arxiv.org/abs/#1}{arXiv:#1}}
\begin{document}
\begin{titlepage}
\begin{flushright}
LPTENS-10-01\\
IFUP-TH/2010-03
\end{flushright}
\vskip 1.0cm
\begin{center}
{\Large \bf Signals of composite electroweak-neutral Dark Matter:
\\[0.3cm] LHC/Direct Detection interplay} \vskip 1.0cm
{\large Riccardo Barbieri$^{a,b}$, Slava Rychkov$^{c}$ and Riccardo Torre$^{b,d}$}\\[0.7cm]
{\it $^a$ Scuola Normale Superiore, Piazza dei Cavalieri 7, I-56126 Pisa, Italy}\\[5mm]
{\it $^b$ INFN, Sezione di Pisa, Largo Fibonacci 3, I-56127 Pisa, Italy}\\[5mm]
{\it $^c$
Laboratoire de Physique Th\'{e}orique, Ecole Normale Superieure,\\
and Facult\'{e} de physique, Universit\'{e} Paris VI, France}\\[5mm]
{\it $^d$
Universit\`a di Pisa, Dipartimento di Fisica, Largo Fibonacci 3, I-56127 Pisa, Italy}
\end{center}
\vskip 1.0cm
\begin{abstract} In a strong-coupling picture of ElectroWeak Symmetry Breaking, a
composite electroweak-neutral state in the TeV mass range, carrying
a global (quasi-)conserved charge, makes a plausible Dark Matter
(DM) candidate, with the ongoing direct DM searches being precisely
sensitive to the expected signals. To exploit the crucial interplay
between direct DM searches and the LHC, we consider a composite
iso-singlet vector $V$, mixed with the hypercharge gauge field, as
the essential mediator of the interaction between the DM particle
and the nucleus. Based on a suitable effective chiral Lagrangian, we
give the expected properties and  production rates of $V$, showing
its possible discovery at the maximal LHC energy with about 100
fb$^{-1}$ of integrated luminosity.
\end{abstract}
\vskip 1cm \hspace{0.7cm} January 2010
\end{titlepage}

\section{Introduction and general properties}

The possibility that Dark Matter (DM) be related, directly or
indirectly, to the physics of ElectroWeak Symmetry Breaking (EWSB)
deserves the highest consideration. Indeed this has been and is
being extensively discussed both in weak-coupling and in
strong-coupling scenarios of EWSB. The strong-coupling case is of
interest to this paper, without specific reference to any detailed
model\footnote{For a review of microscopic models of strong EWSB see
\cite{HS}.}.

We consider the case where the forces responsible for EWSB respect a global
(quasi-)conserved charge $X$ which enforces the (quasi-)stability of the
lightest particle, $\Phi$, with non vanishing $X$. $\Phi$ is a candidate DM
particle. Its mass, $m_{\Phi}$, is in the TeV range, characteristic of the
strong forces that may give rise to EWSB. This particle is made of
constituents that feel the strong force and carry non-vanishing electroweak
quantum numbers, but is itself electroweak-neutral. This is needed in order to
suppress the tree-level coupling of $\Phi$ to the Z-boson, which would be in
conflict with direct DM searches.

At face value, the cosmological relic abundance of the $\Phi$-particles is too
low to explain the observed DM energy density, $\Omega_{DM}$, normalized as
usual to the critical cosmological density. We have in mind the effect of two
body processes, $\Phi\bar{\Phi}\leftrightarrow Q\bar{Q}$, where $Q$ is any
unstable particle lighter than $\Phi$, also feeling the new strong force. For
example, longitudinal W and Z bosons may play the role of $Q$. The associated
thermally averaged cross section is far bigger than the needed $\left\langle
\sigma v\right\rangle \approx1$pb, since
\begin{equation}
\left\langle \sigma v\right\rangle \approx\frac{\lambda^{4}}{4\pi m_{\Phi}%
^{2}}f\left(  \frac{m_{\Phi}^{2}}{\Lambda^{2}}\right)  \approx10^{6}%
\,\text{pb}\left(  \frac{\lambda}{4\pi}\right)  ^{4}\left(  \frac{\text{TeV}%
}{m_{\Phi}}\right)  ^{2}f\left(  \frac{m_{\Phi}^{2}}{\Lambda^{2}}\right)  \,,
\end{equation}
where ${\lambda}\approx{4\pi}$ is a Naive Dimensional Analysis (NDA) estimate
of the strong coupling ${\lambda}$, and the model-dependent function $f$ of
the ratio between the $\Phi$-mass and the scale $\Lambda$ characteristic of
the new strong interaction is of order unity for $m_{\Phi}\approx\Lambda
$.\footnote{In Ref.~\cite{Nomura}, strongly interacting DM belonging to a
non-EWSB hidden sector was considered, with a thermal freezout as the source
of DM abundance. However, much higher DM\ masses up to 100 TeV were considered
and smaller than NDA couplings were assumed.}

We are thus led to consider the case where the relic abundance of $\Phi
$-particles originates from a $X-\bar{X}$ asymmetry, analogous to the standard
$B-\bar{B}$ asymmetry responsible for the dominance of matter over anti-matter
in the present universe. An interesting aspect of this hypothesis is that one
can try to relate $\Omega_{X}=\Omega_{DM}$ to the standard $\Omega_{B}$ by
assuming that $X$, like $B$ or $L$, are all broken by mixed electroweak
anomalies. In this case, non-perturbative electroweak sphaleron interactions
at a critical temperature $T^{\ast}\approx100\div200$ GeV may redistribute any
original asymmetry, leading in particular today to \cite{Kaplan}
\begin{equation}
\frac{\Omega_{DM}}{\Omega_{B}}=\mathcal{O}(10^{2})x^{5/2}e^{-x},~x=\frac
{m_{\Phi}}{T^{\ast}},
\end{equation}
which can be about right, ${\Omega_{DM}}/{\Omega_{B}}\approx5$, for $m_{\Phi}$
in the TeV range.

In this as in other cases of putative DM particles, the problem is to find
experimental signals that would not only establish their existence but would
allow a clear interpretation of their nature. To this end it is difficult to
overestimate the interplay between direct DM searches and LHC experiments, as
we are going to discuss.

\section{Summary of direct detection signals}

In absence of a detailed model, the possible signal in direct detection
searches can be discussed by means of effective operators that mediate the
interaction between $\Phi$ and the $u,d$-quarks or the photon \cite{Bagnasco}.
The $\Phi$-particle can be either a complex scalar or a Dirac fermion.

If $\Phi$ is a scalar, the dominant interactions are described by
\begin{equation}
O_{1}=\frac{1}{\Lambda^{2}}(\Phi^{\ast}\overleftrightarrow{\partial_{\mu}}%
\Phi)\sum_{q=u,d}c_{q}(\bar{q}\gamma_{\mu}q),\qquad O_{2}=\frac{ec_{2}%
}{\Lambda^{2}}\partial_{\mu}F^{\mu\nu}(\Phi^{\ast}\overleftrightarrow
{\partial_{\mu}}\Phi)\,, \label{operators}%
\end{equation}
which for $c_{q}\approx c_{2}\approx1$, as expected from NDA, give comparable
effects. Taking $O_{2}$ for concreteness, the non-relativistic cross section
of $\Phi$ on a nucleus of charge $Z$ and mass $m_{N}\ll m_{\Phi}$ is, up to
form factor effects,
\begin{equation}
\quad\sigma_{2}=c_{2}^{2}\frac{e^{4}Z^{2}m_{N}^{2}}{\pi\Lambda^{4}}\,.
\end{equation}
For Germanium target this corresponds to the per-nucleon cross section%
\begin{equation}
\frac{\sigma_{2}}{A^{4}}\approx c_{2}^{2}\,2\cdot10^{-7}\text{pb}\left(
\frac{\text{TeV}}{\Lambda}\right)  ^{4}, \label{eq:scalar}%
\end{equation}
to be compared with the CDMS limit on the coherent spin-independent cross
section \cite{Ahmed:2009zw}
\begin{equation}
\frac{\sigma_{\text{SI}}}{A^{4}}|_{\exp}\lesssim2\cdot10^{-7}\text{pb}\left(
\frac{m_{DM}}{\text{TeV}}\right)  \qquad(m_{DM}\gg m_{\text{Ge}}),
\end{equation}
i.e.
\begin{equation}
c_{2}<\left(  \frac{\Lambda}{\text{TeV}}\right)  ^{2}\left(  \frac{m_{\Phi}%
}{\text{TeV}}\right)  ^{1/2}. \label{bound}%
\end{equation}
For $c_{2}\approx1,$ and $\Lambda\approx m_{\Phi}\approx4\pi v\approx3$ TeV,
where $v\approx250$ GeV is the electroweak VEV, the expected cross section
(\ref{eq:scalar}) is about two orders of magnitude below the CDMS limit.

If instead $\Phi$ is a Dirac fermion, assuming parity invariance (up to
anomalies) of the EWSB forces, the dominant operator is a magnetic moment
interaction
\begin{equation}
O_{M}=\frac{iec_{M}}{2\Lambda}(\bar{\Phi}\sigma_{\mu\nu}\Phi)F^{\mu\nu}.
\end{equation}
In Germanium, $O_{M}$ gives rise to the dominant spin-independent cross
section due to scattering on the current produced by the nuclear charge:
\begin{equation}
\frac{d\sigma_{M}}{dE}\approx c_{M}^{2}\frac{e^{4}Z^{2}}{4\pi\Lambda^{2}%
E}(1+\mathcal{O}(E/E_{\text{max}})), \label{eq:fermi-si}%
\end{equation}
where $E$ is the kinetic energy of the recoiling nucleous, ranging from the
experimental threshold $\sim10$ keV to $E_{\max}=2v^{2}m_{N}^{2}%
=\mathcal{O}(100$ keV). The spin-dependent cross section is subleading due to
small nuclear spin of Germanium and because of $E_{\max}/E$ enhancement
present in (\ref{eq:fermi-si}) \cite{Bagnasco}. A suitable comparison of this
cross section with the null CDMS result gives in this case
\begin{equation}
c_{M}<10^{-1}\left(  \frac{\Lambda}{\text{TeV}}\right)  \left(  \frac{m_{\Phi
}}{\text{TeV}}\right)  ^{1/2},
\end{equation}
against the NDA estimate $c_{M}\approx1$. Given the uncertainties of these
estimates and of the value of the scale $\Lambda$ itself, in no way this bound
can be interpreted as ruling out a composite fermionic DM particle. Quite on
the contrary, the message we draw is that a signal in direct DM searches could
be around the corner. Yet we find it preferable, at least for reference, to
stick in the following to the scalar case.

\section{The DM-nucleus interaction mediated by a vector iso-singlet $V$}

\begin{figure}[ptbh]
\centering
\includegraphics[width=4cm]{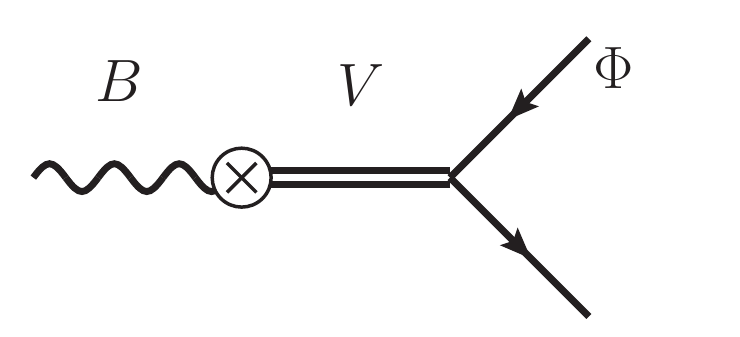} \caption{The diagram which generates
$O_{2}$ via the $V-B$ mixing.}%
\label{VB}%
\end{figure}

Suppose that a positive signal were indeed found in direct DM searches at the
level indicated above, in fact not far from the present sensitivity. How would
we know that the candidate DM particle is a composite $\Phi$-like particle? As
already mentioned, LHC should come into play here. However the detection at
LHC of an electroweak-neutral particle of TeV mass that can only be pair
produced may not be an easy task. For this reason, we turn the question into a
different but related one. What could mediate the operators in Eq.
(\ref{operators}) responsible in the first place for the direct DM signal? We
argue that the most likely candidate for this role is a composite vector
iso-singlet $V$, the analog of the $\omega$-meson in QCD, strongly coupled to
$\Phi$ and mixed with the elementary hypercharge gauge boson $B_{\mu}$, via
the diagram of Fig.\ \ref{VB}.

We base our estimates on the following phenomenological Lagrangian
\begin{equation}
\mathcal{L}=\mathcal{L}_{V}+\mathcal{L}_{V\Phi} \label{Ltot}%
\end{equation}
where
\begin{equation}
\mathcal{L}_{V\Phi}=g_{S}V_{\mu}(\Phi^{\ast}\overleftrightarrow{\partial_{\mu
}}\Phi),\quad g_{S}=4\pi\frac{M_{V}}{\Lambda}, \label{gS}%
\end{equation}
and
\begin{equation}
\mathcal{L}_{V}=-\frac{1}{4}V_{\mu\nu}^{2}+\frac{1}{2}M_{V}^{2}V_{\mu}%
^{2}+\frac{g^{\prime}}{4\pi}B_{\mu\nu}V_{\mu\nu}-\frac{i}{8\pi}\epsilon
^{\mu\nu\rho\sigma}V_{\mu}\text{tr(}u_{\nu}u_{\rho}u_{\sigma}\text{)}+\frac
{g}{4\pi}\epsilon^{\mu\nu\rho\sigma}V_{\mu}\text{tr(}u_{\nu}\hat{W}%
_{\rho\sigma}\text{)} \label{lagrangian}%
\end{equation}
in the standard notation for the electroweak chiral Lagrangian, i.e.
\begin{equation}
u_{\mu}=iuD_{\mu}Uu^{+}\approx i\partial_{\mu}U+g^{\prime}B_{\mu}\frac
{\sigma_{3}}{2}-gW_{\mu}^{a}\frac{\sigma_{a}}{2},\quad U=u^{2}=e^{(i\sigma
_{a}\pi^{a}/v)},
\end{equation}
$\hat{W}_{\mu\nu}=W_{\mu\nu}^{a}\sigma_{a}/2$ is the usual field strength for
the W boson, and the $\pi$-fields are the eaten up Goldstone bosons for EWSB.
We assume that the couplings proportional to the epsilon tensor, relevant to
the following Section, are induced, analogously to the QCD case, by a chiral
anomaly. The strength of the various couplings are all based on NDA estimates,
known to work well for QCD \cite{Klingl}, and noticing that the only coupling
in (\ref{Ltot}) that corrects at one loop level the $V$-mass $M_{V}$ is
$g_{S}$ in (\ref{gS}).

From the diagram of Fig.\ \ref{VB} it is straightforward to obtain the
operator $O_{2}$ in eq. (\ref{operators}) with
\begin{equation}
c_{2}=\frac{2\Lambda}{M_{V}},
\end{equation}
or, from (\ref{bound}),
\begin{equation}
M_{V}>2\,\text{TeV}\left(  \frac{\text{TeV}}{\Lambda}\right)  \left(
\frac{\text{TeV}}{m_{\Phi}}\right)  ^{1/2},
\end{equation}
which could easily allow, taking $\Lambda\approx m_{\Phi}\approx4\pi
v\approx3$ TeV, a $V$-mass as low as 700 GeV. The iso-singlet nature of $V$
makes its exchange innocuous in the ElectroWeak Precision Tests, giving a
contribution to the $Y$-parameter \cite{LEPII} well below the $10^{-4}$ level.

\section{LHC phenomenology of $V$}

The Lagrangian (\ref{lagrangian}) allows to calculate the decay widths of $V$.
The relevant widths are:

\begin{itemize}
\item From the last term in (\ref{lagrangian}), the dominant decay into two
standard vector bosons\footnote{This formula corrects a factor 9 error in the
RHS of Eq.~(4.7) of \cite{Klingl}.}%
\begin{gather}
\Gamma_{\text{tot}}\approx\Gamma\left(  V\rightarrow W^{+}W^{-},ZZ,Z\gamma
\right)  =\frac{g^{2}}{8\pi}\frac{M_{V}^{3}}{(4\pi v)^{2}}\,,\\
\text{BR}\left(  V\rightarrow W^{+}W^{-}\right)  \approx\frac{2}{3}%
,\qquad\text{BR}\left(  V\rightarrow ZZ\right)  \approx\frac{\cos^{2}%
\theta_{W}}{3},\qquad\text{BR}\left(  V\rightarrow Z\gamma\right)
\approx\frac{\sin^{2}\theta_{W}}{3}\,.
\end{gather}

\item From the last but one term in (\ref{lagrangian}), the 3-body decay%
\begin{equation}
\Gamma\left(  V\rightarrow W_{L}^{+}W_{L}^{-}Z_{L}\right)  =\frac{\pi}%
{40}\frac{M_{V}^{7}}{(4\pi v)^{6}}\,.
\end{equation}

\item From the mixing of $V$ with the $B$ boson, the decay into a pair of
standard fermions, e.g.
\begin{equation}
\Gamma\left(  V\rightarrow e^{+}e^{-}\right)  =\frac{5}{24\pi}\left(
\frac{g^{\prime2}}{4\pi}\right)  ^{2}M_{V}\ .%
\end{equation}

\end{itemize}

\begin{figure}[ptbh]
\begin{minipage}[b]{8.2cm}
\centering
\includegraphics[width=8.5cm]{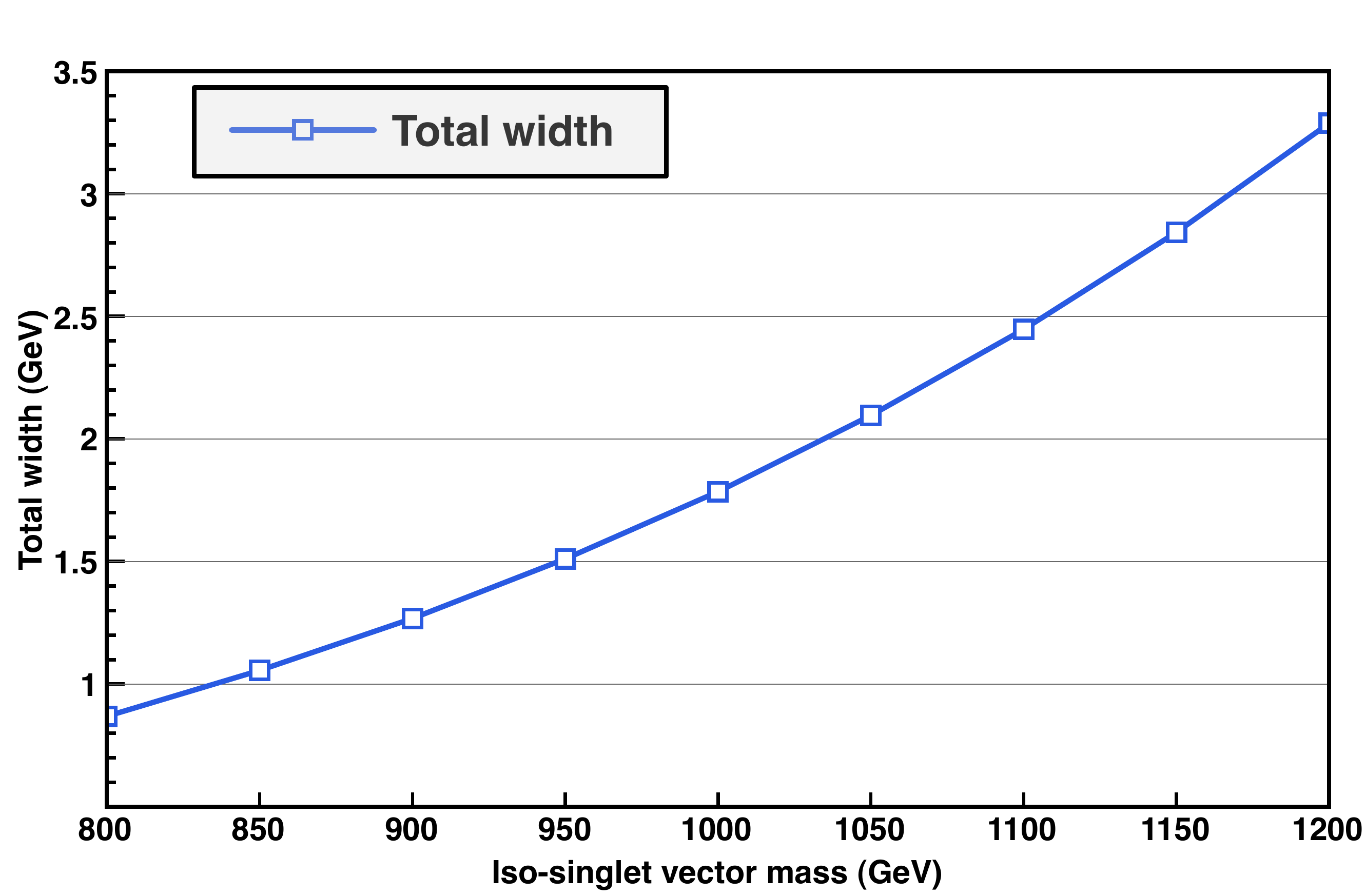}
\end{minipage}
\ \hspace{2mm} \hspace{3mm} \ \begin{minipage}[b]{8.5cm}
\centering
\includegraphics[width=8.5cm]{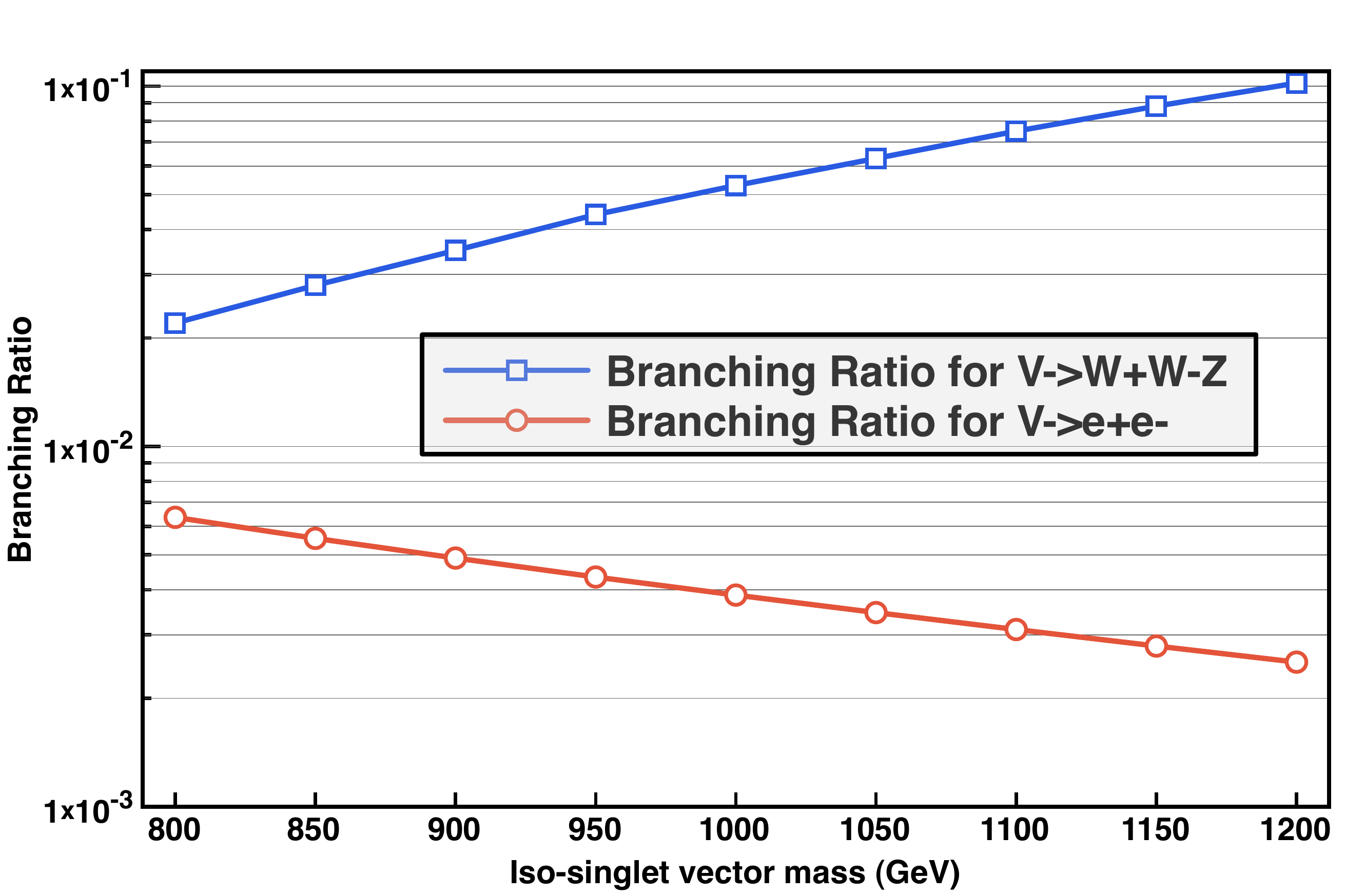}
\end{minipage}
\caption{Left panel: the total width of the iso-singlet vector boson $V$ as a
function of its mass around $1$ TeV. Right panel: the subdominant branching
ratios $BR(V\to e^{+}e^{-})$ and $BR(V\to W^{+}W^{-}Z)$.}%
\label{width_BR}%
\end{figure}

The total width and the subdominant branching ratios are shown in
Fig.\ \ref{width_BR} for $M_{V}$ around 1 TeV. A few features are especially
apparent from these figure: the smallness of the $\Gamma/M$ ratio and the
strong dominance of the decays into two bosons (among which the $Z\gamma$
channel) over all the other decay modes, in particular the three body
$W^{+}W^{-}Z$. Especially this last feature is at variance with what one might
have expected from the analogy with the $\omega$ in QCD. The main reason for
this can be traced back to the close degeneracy of the $\omega$ with the
$\rho$, as dictated by $SU(3)$, making the decay $\omega\rightarrow3\pi$
dominated by the intermediate $\pi\rho$ state.

\begin{figure}[ptbh]
\begin{center}
\includegraphics[width=12cm]{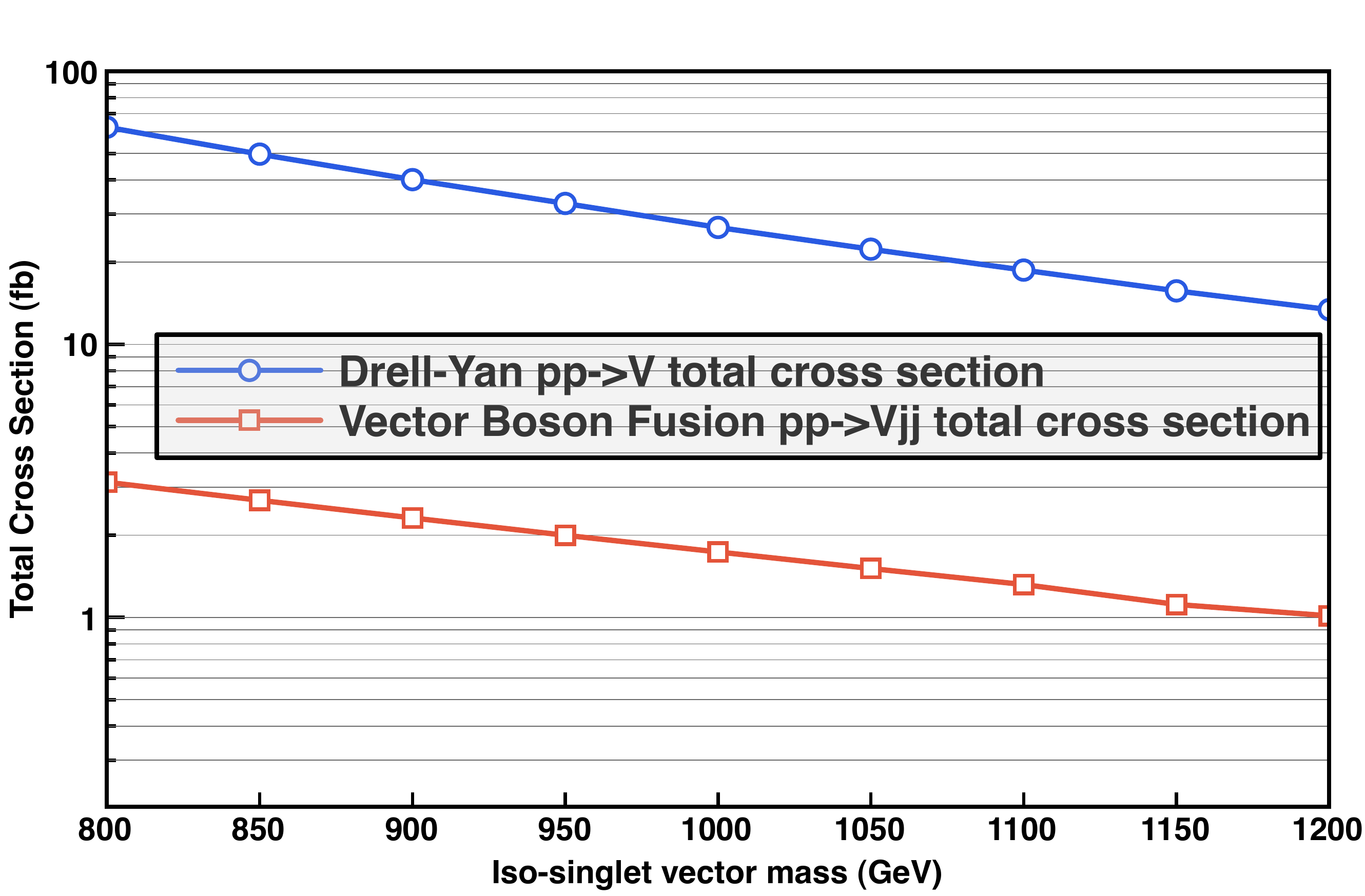}
\end{center}
\caption{Total cross sections for the Vector Boson Fusion and the Drell-Yan
iso-singlet vector boson production at the LHC as funcions of its mass for
$\sqrt{s}=14$ TeV.}%
\label{totalcrosssections}%
\end{figure}

\begin{figure}[ptbh]
\begin{minipage}[b]{8.2cm}
\centering
\includegraphics[width=8.5cm]{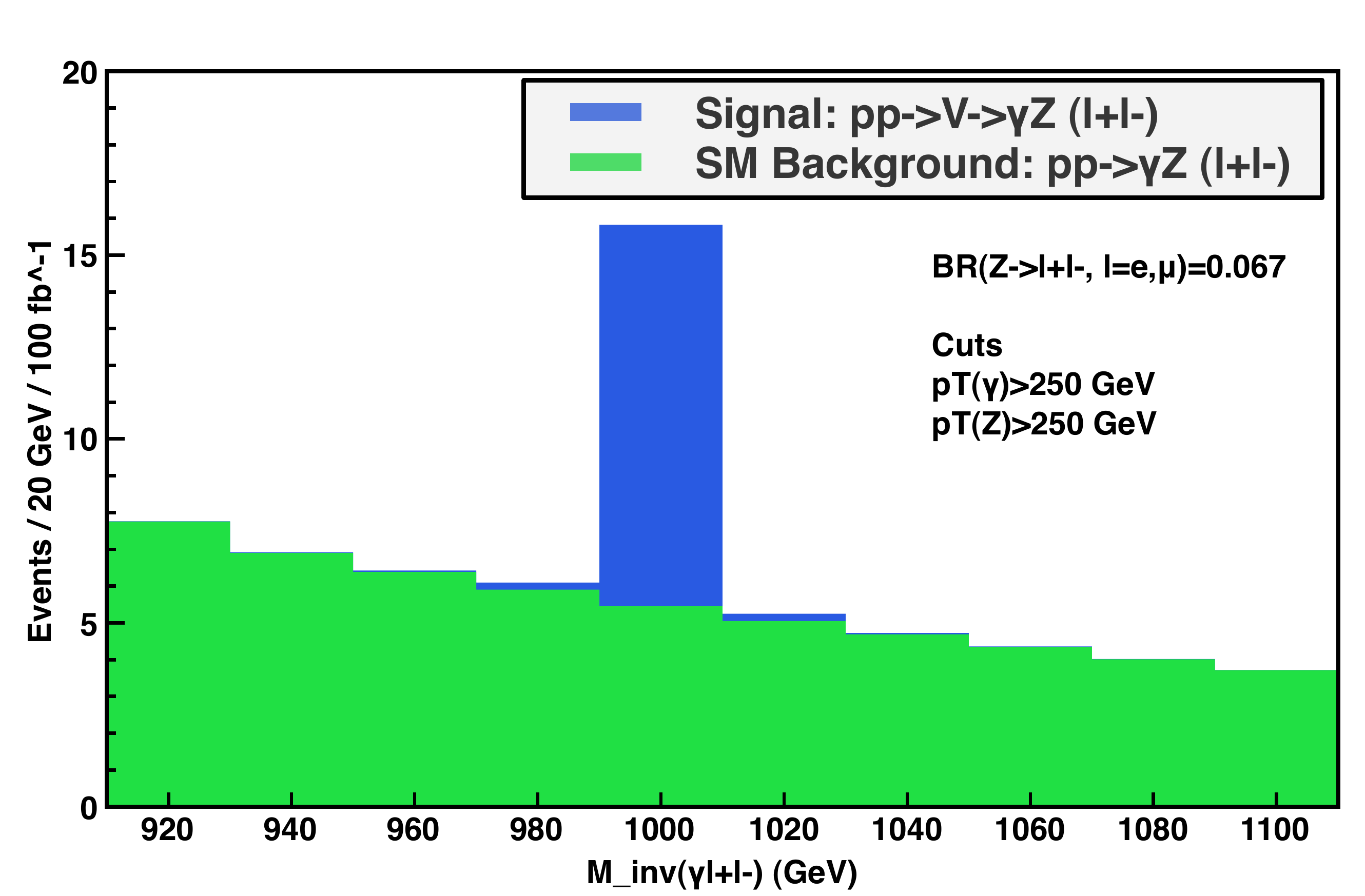}
\end{minipage}
\ \hspace{2mm} \hspace{3mm} \ \begin{minipage}[b]{8.5cm}
\centering
\includegraphics[width=8.5 cm]{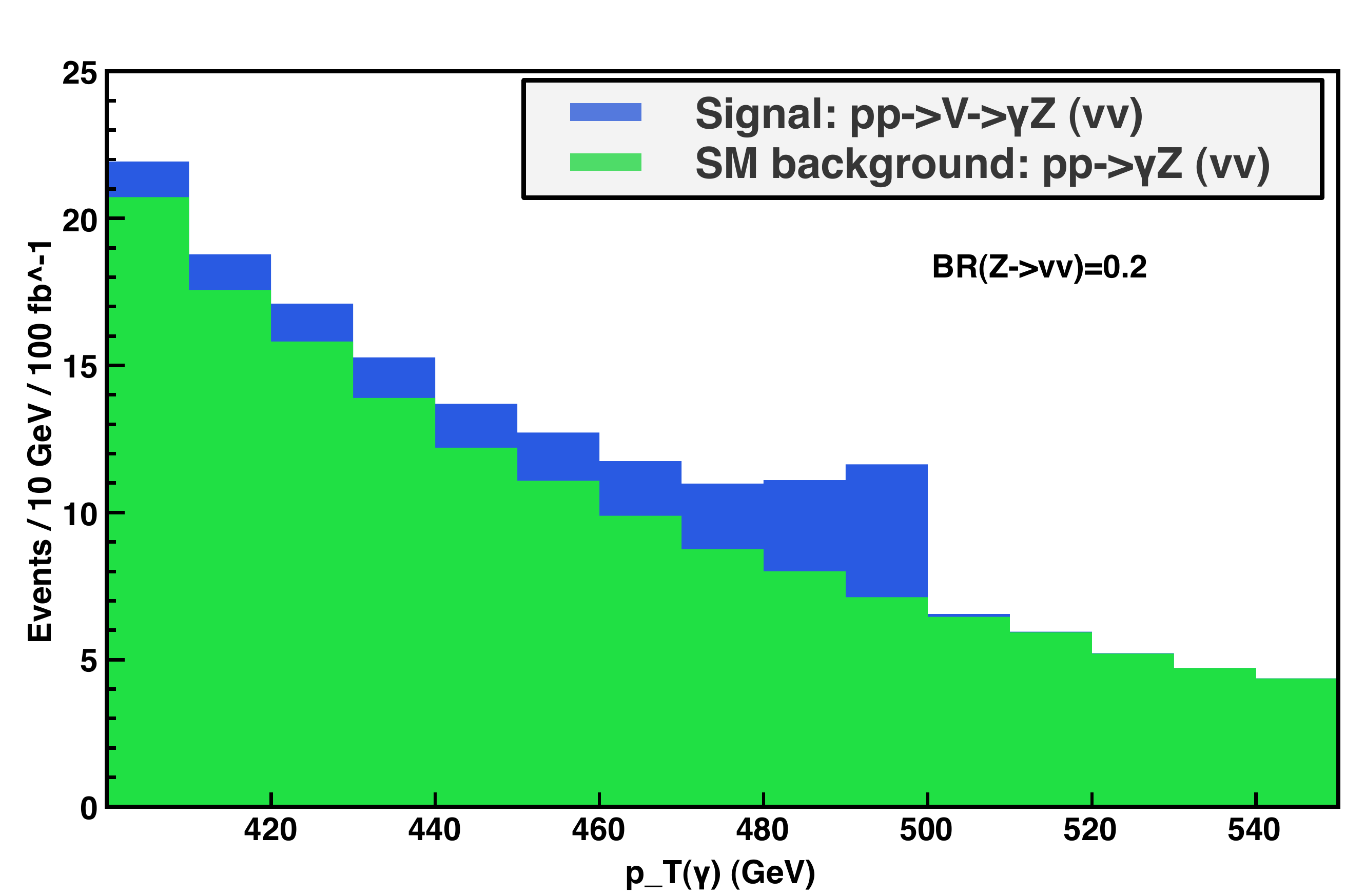}
\end{minipage}
\caption{Left panel: the number of $\gamma l^{+}l^{-}$ events as a
function of the total invariant mass. The imposed $p_{T}>250$ GeV
cut on the photon and the reconstructed $Z$ boson enhances the $S/B$
ratio. A $\sim5\sigma$ excess from the SM prediction can be seen.
Right panel: the number of $\gamma+\slash\!\!\!\!{E}_{T}$ events as
a function of the photon transverse momentum. This decay mode will
give additional information although not a discovery
($S/\sqrt{B}\sim 2$). Both plots are for 100 fb${}^{-1}$ of data at
$\sqrt{s}=14$ TeV. The
chosen binnings correspond to twice the expected experimental resolution. }%
\label{AZ}%
\end{figure}

The vector $V$ can be produced at the LHC by the Drell-Yan (DY) process or by
Vector Boson Fusion (VBF), as again described by the Lagrangian
(\ref{lagrangian}). The corresponding production rates are shown in Fig.
\ref{totalcrosssections}. 
From the branching ratios  above, the $Z\gamma$ channel appears most promising. In Fig.\ \ref*{AZ}
we show the number of events expected when the $Z$ decays into $l^{+}%
l^{-},l=e,\mu$ or into $\nu\bar{\nu}$ respectively at $\sqrt{s}=14$
TeV. The binnings of the events, crucial for discovery, are based on
current estimates of the expected resolutions in an advanced phase
of LHC operation\footnote{We are assuming a 1\% energy resolution of
the invariant mass of the $l^{+}l^{-}\gamma$ system (comparable to
the peak resolution in the $H\to\gamma\gamma$ studies) and a $0.5\%$
resolution of the photon $p_T$. The binnings correspond to $2\sigma$
bins.} The background shown corresponds to the $Z\gamma$ production
in the Standard Model. On the basis of these figures, we conclude
that the vector $V$ in the TeV mass range could be discovered at LHC
with about 100 fb$^{-1}$ of integrated luminosity.\footnote{See
\cite{Tevatron} for a recent D$\not\!0$ search of narrow vector
resonances decaying into $Z\gamma$ based on 1 fb${}^{-1}$ of
Tevatron data. The resulting limit on $\sigma\times$B.R. is
$\sim0.2-0.4$ pb (95\% C.L.) for $M_V=700-900$ GeV, two orders of
magnitude above the values predicted by our model.}


\section{Summary and conclusions}

The strong-coupling scenarios of EWSB deserve attention in spite of generic
difficulties in satisfying the ElectroWeak Precision Tests. Due to the lack of
calculability, one is forced to use a phenomenological Lagrangian description
of the low-lying resonances with coefficients of various operators estimated
from NDA. This philosophy has been often applied in the studies of the
isospin-1 resonances (`techni-$\rho$') \cite{rho}. In this paper, we used it
to study another interesting generic corner of the strong sector, consisting
of an iso-singlet vector $V$ (`techni-$\omega$') coupled to the lightest
particle $\Phi$ carrying nonvanishing conserved charge (`techni-baryon
number'). The $\Phi$ is a candidate DM particle, assumed electroweak-neutral
to evade direct detection constraints\footnote{For the alternative possibility
of a neutral iso-triplet component, see \cite{Frandsen:2009mi}.}.

We have shown that the phenomenology of this sector allows for an interesting
interplay between the ongoing direct DM searches and the LHC. Apart from NDA,
our main assumption is that the operators describing interactions of $\Phi$
with the SM particles, Eq.~(\ref{operators}), are generated via the cubic
$\Phi\Phi V$ coupling in the strong sector, and the mixing between $V$ and the
elementary hypercharge gauge field $B,$ Fig.~\ref{VB}. This is the Vector
Meson Dominance hypothesis, which is known to work well in QCD not only for
the pion \cite{Klingl} but also for the nucleons \cite{VMD}. Two conclusions
transpire from our analysis. First, the expected signal in direct detection
experiments, estimated already in \cite{Bagnasco}, is not far from the present
experimental bounds. Second, the isosinglet vector $V$, in its typical mass
range and with couplings as in (\ref{lagrangian}), appears within reach of the LHC with $\mathcal{O}(100$fb$^{-1})$ integrated
luminosity\footnote{For an old study of the techni-$\omega$ discovery
potentials at the SSC see \cite{Chivu}. That study correctly identifies the
crucial $Z\gamma$ decay mode. Our approach to estimating the production cross
section is more direct, in our opinion.}. Its strong sector nature will be
easily identifiable due to a characteristic $Z\gamma$ decay mode. The joint
observation of nuclear recoil events in direct detection experiments and of a
vectorial resonance decaying into $Z\gamma$ at the LHC will then point to a
composite DM-particle.

\subsection*{Acknowledgements}

We~are grateful to M.~Mangano and G. Rolandi for useful
communications concerning the binnings in Fig.\ \ref{AZ}. This
research was supported in part by the European Programme
``Unification in the LHC Era", contract PITN-GA-2009-237920
(UNILHC).

\end{document}